\documentclass[reprint, amssymb, amsmath, prl, cha, superscriptaddress]{revtex4-1}

\usepackage{graphicx}

\usepackage[colorlinks=true,linkcolor=blue]{hyperref}
\expandafter\ifx\csname package@font\endcsname\relax\else
 \expandafter\expandafter
 \expandafter\usepackage
 \expandafter\expandafter
 \expandafter{\csname package@font\endcsname}
\fi
\begin{document}


\title{Sampling polymorphs of ionic solids using random superlattices}



\author{Vladan Stevanovi\'c}
\email{vstevano@mines.edu}
\affiliation{Colorado School of Mines, Golden, Colorado 80401, USA}%
\affiliation{National Renewable Energy Laboratory, Golden, Colorado 80401, USA}%
\date{\today}

\begin{abstract}
Polymorphism offers rich and virtually unexplored space for discovering novel functional materials.
To harness this potential approaches capable of both exploring the space of
polymorphs and assessing their {\it realizability} are needed.  One such approach devised for partially 
ionic solids is presented. The structure prediction part is carried out by performing local DFT relaxations 
on a large set of random supperlattices (RSLs) with atoms distributed randomly over different planes in a 
way that favors cation-anion coordination. Applying the RSL sampling on MgO, ZnO and SnO$_2$ reveals 
that the resulting probability of occurrence of a given structure offers a measure of its realizability explaining 
fully the experimentally observed, metastable polymorphs in these three systems.
\end{abstract}
%
%
\maketitle
%
%
The discovery of polymorphism in the late 18th and early 19th century \cite{thernard:1809, verma_polymorphs:1966}
revealed the significance of structural degrees of freedom in determining physical properties of solids.
The best known example is probably elemental carbon with markedly different mechanical, optical and electronic 
properties between its graphite and diamond forms \cite{LBDB}. Other notable cases include white and grey 
tin, which also exhibit significant differences in electronic and mechanical properties \cite{LBDB}; or enhanced 
photocatalytic activity of anatase TiO$_2$ compared to the ground state rutile polymorph \cite{luttrell_SciRep:2014}; 
or elemental Silicon, an indirect band-gap semiconductor in the ground state diamond structure predicted to become 
a direct gap material in a number of higher energy structures \cite{botti_PRB:2012} including the experimentally realized 
clathrate structure \cite{baranowski_JMCC:2014}. 

However, the development of rational approaches to explore the space of polymorphs and (desirably) assist in their experimental 
realization faces significant challenges. First, the complexity of the potential energy surface (PES) of periodic systems, evidenced by the 
exponential increase in the number of local minima with the system size \cite{stillinger_PRE:1999}, limits our ability to systematically 
explore the spectrum of possible structures. 

A related problem of finding the ground state structure attracted 
attention, especially with the development of first-principles total energy methods, resulting in a number of 
structure prediction techniques \cite{catlow_NMAT:2008}. These include simulated annealing \cite{pannetier_Nat:1990, schon_ACIEd:1996}, 
methods based on evolutionary algorithms \cite{woodley_PCCP:1999, oganov_JPC:2006, trimarchi_PRB:2007}, metadynamics 
\cite{laio_PNAS:2002, barducci_CMS:2011}, basin and minima hopping \cite{wales_JPCA:1997,goedecker_JCP:2004},  random structure 
searching \cite{pickard_JPCM:2011}, methods based on data mining and machine learning \cite{fischer_NMAT:2006}, structure prototyping 
\cite{stevanovic_PRB:2012, gautier_NMAT:2015}, etc. Although focused on finding the ground state structure some of these 
methods were also used in exploring the space of polymorphs (see for example  Refs.~\cite{botti_PRB:2012,huan_PRL:2013}) with the energy 
above the ground state as the main quantifier of their potential for experimental realization.

This brings us to the second major challenge, the assessment of the likelihood for experimental realization of 
different polymorphs. While certainly being an important quantity, the energy above the ground state alone
is insufficient to explain observations based on available experimental data. For example, in the case of MgO, despite predictions \cite{zwijnenburg_PRB:2011} 
only the ground state rocksalt structure, and no other, is experimentally realized as reported in the Inorganic Crystal Structure Database \cite{icsd}. 
In the case of ZnO, only the ground-state wurtzite and two other structures, 
zincblende and rocksalt, are experimentally realized \cite{icsd, ashrafi_APL:2000, decremps_APL:2002}. 
Another important example is SnO$_2$, which undergoes a series of phase transitions under pressure \cite{shieh_PRB:2006}, but all of the 
high pressure phases relax to either the ground state rutile or the metastable $\alpha$-PbO$_2$ structure type 
upon releasing the pressure \cite{liu_Science:1978}. These facts indicate that for a given composition there seems to exist a finite set of structures 
that have higher likelihood for realization than the rest.
%
\begin{figure}[t!]
\includegraphics[width=\linewidth]{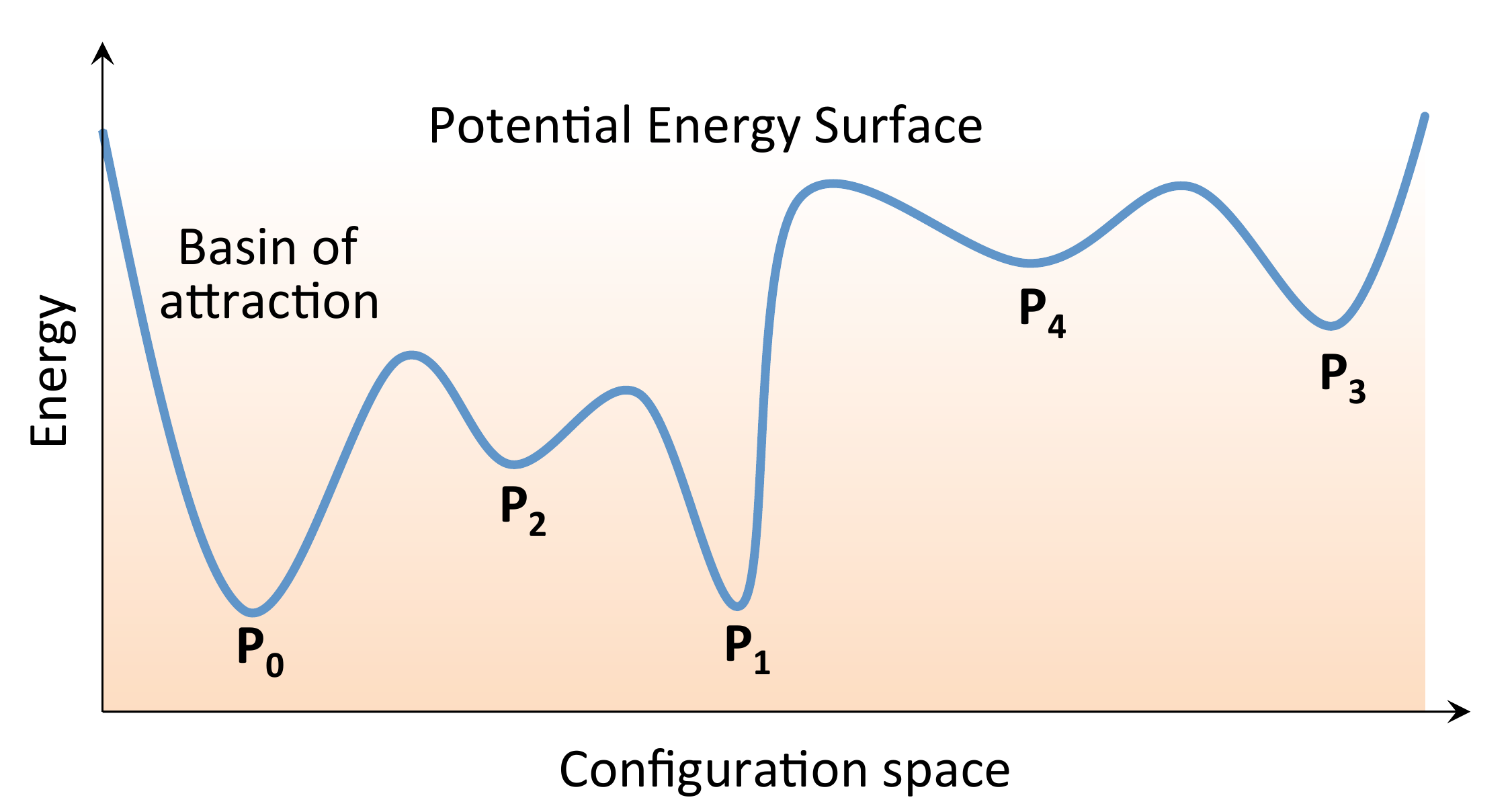}
\caption{\label{fig1} (color online) A sketch of the potential energy surface (PES) of solids with different polymorphs corresponding to
different PES local minima. }
\end{figure}
%
%
\begin{figure*}[t!]
\includegraphics[width=0.75\linewidth]{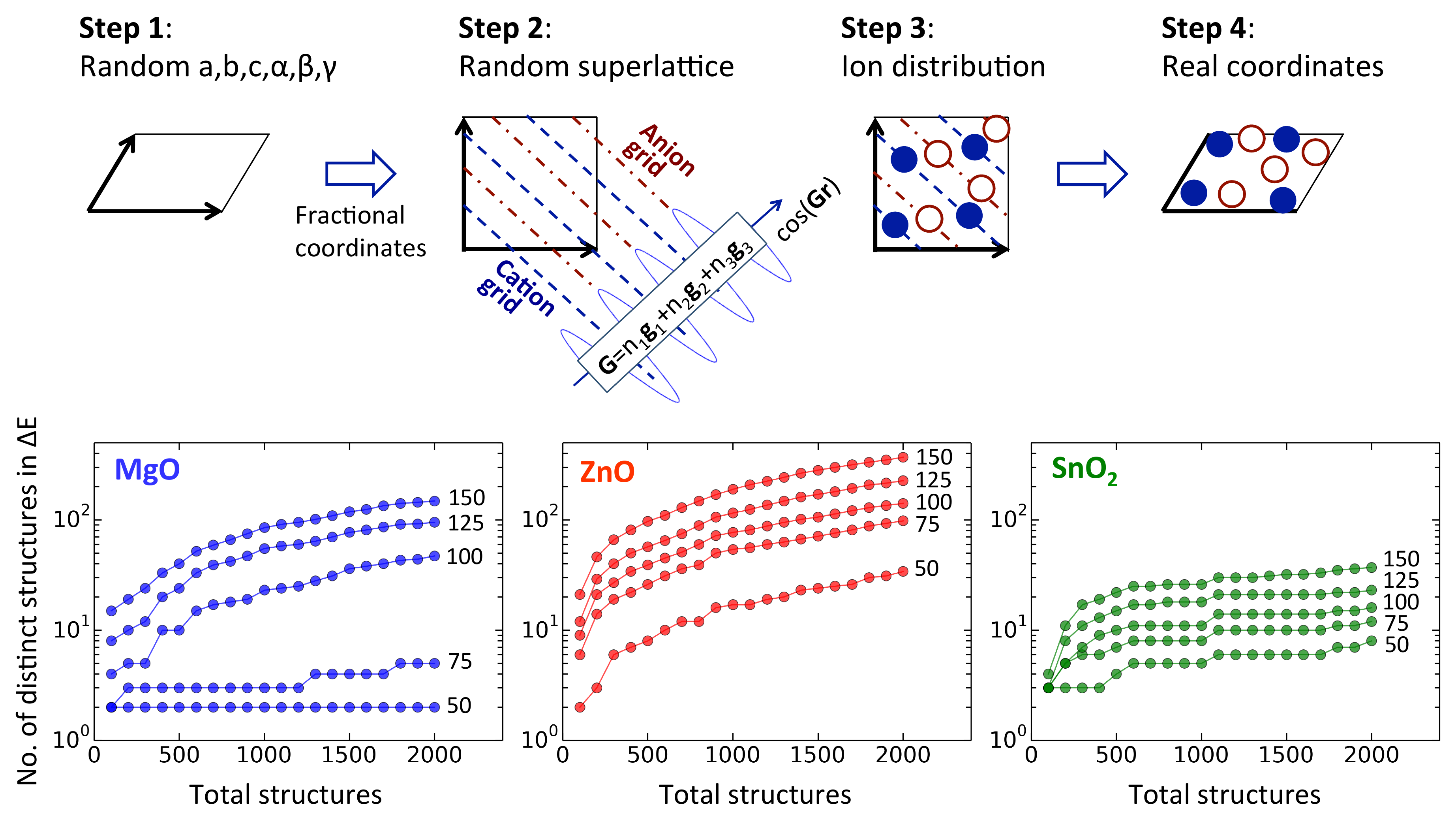}
\caption{\label{fig2} (color online) Steps in the random superlattice (RSL) structure generation (upper panel).  Lower panel shows the
number of distinct structures (polymorphs) within $\Delta E= 50, 75, 100, 125, 150$ meV/atom above the ground state as a function of the total 
number of structures used in the sampling procedure.}
\end{figure*}

If this is true, then the realizability of a given polymorph can be thought of as determined by a  combination of three factors: 
({\it i}) the energy above the ground state, 
({\it ii}) the energy barrier to escape from a given PES minimum, and 
({\it iii}) the volume of configuration space occupied by the PES minimum.  
The ({\it i}) and ({\it ii}) describe the principles of energy minimization and kinetic trapping. The factor ({\it iii}) on the other hand, 
measures the probability of getting into a given PES minimum. More precisely, as shown in Fig.~\ref{fig1} every local minimum on 
the PES defines its basin of attraction, or the region of configuration space that has that minimum as its "center of gravity". Hence, the 
probability of "falling" into a certain structure has to be proportional to the total volume of configuration space occupied by its basin of 
attraction including all of the symmetry equivalent basins. 
%
\begin{figure*}[t!]
\includegraphics[width=0.85\linewidth]{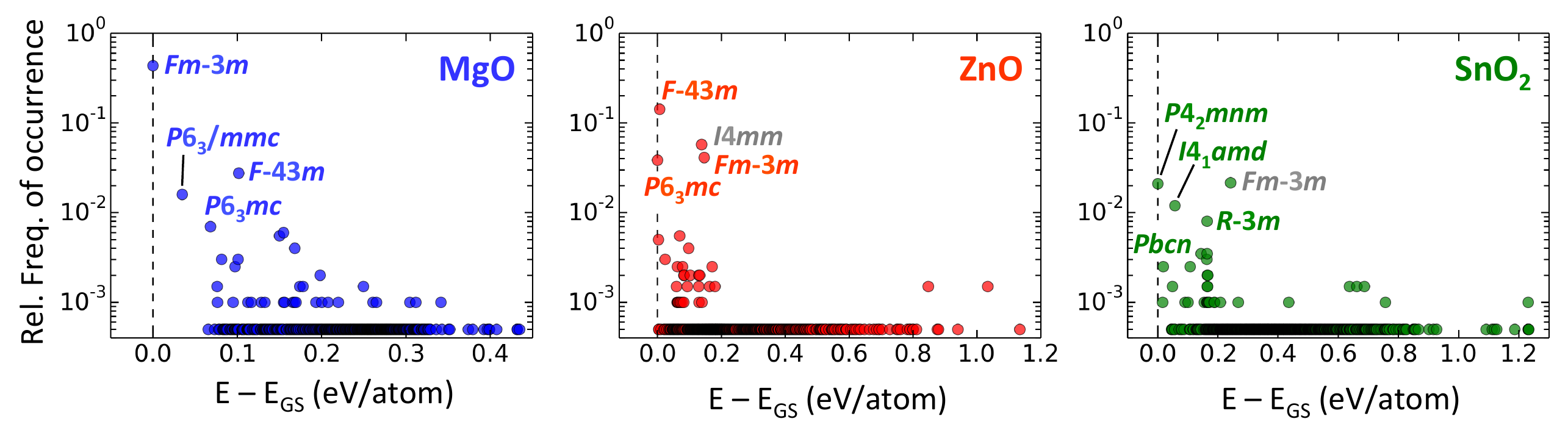}
\caption{\label{fig3} (color online) Relative frequencies of occurrence of structures resulting from the RSL sampling shown 
against the energy above the grounds state. Only the top occurring structures have their space groups explicitly marked. Space groups shown 
in grey represent the structures that are predicted to be the dynamically unstable.}
\end{figure*}

Herein it is demonstrated, using MgO, ZnO and SnO$_2$ as case examples, that the factor ({\it iii}) is actually critical in establishing a 
ranking of realizability with ({\it i}) and ({\it ii}) providing additional constraints. This is done by pursuing the idea that the total volume 
occupied by various basins of attraction can be measured using a large number of random structures (random unit cell vectors and 
random atomic positions) that are relaxed to the closest PES local minimum utilizing density functional theory (DFT). 
The frequency of occurrence of a given structure would then provide a measure of the probability to "fall" into its local minimum.

To do this and, at the same time, to overcome in part the difficulties posed by the already mentioned complexity of the PES, a structure prediction 
method is proposed to bias the random sampling toward the region of the PES more relevant for ionic systems. Because of the charge transfer 
only the structures that have cations preferentially coordinated by anions and vice-versa are relevant. The method adopted here 
favors the cation-anion coordination by distributing different types of ions in a random fashion over two interpenetrating grids of points. 
The grids are constructed using the alternating planes of a superlattice defined by a randomly chosen reciprocal lattice vector (see Fig.~\ref{fig2}). 
Constructed in this way, these random superlattice (RSL) structures exhibit dominant cation-anion coordination.

For each MgO, ZnO, and SnO$_2$ a total of 2000 RSL structures with sizes varying 
between 1--20 formula units are constructed and DFT-relaxed to the closest local minimum. The relaxed structures are sorted into classes of 
equivalence and for the classes with largest occupancies (frequencies of occurrence) additional phonon calculations are performed
with the purpose of providing the information on the dynamic stability. The analysis of the resulting 
frequencies of occurrence shown in Fig.~\ref{fig3}
reveals that the experimentally observed polymorphs are exclusively the ones with the highest occurrence. 

Namely, for MgO the relative frequency of occurrence of the rocksalt phase is $\sim$0.44, and is more than an order of magnitude larger than for any 
other structure, which explains why is the rocksalt phase so ubiquitous in MgO.  Experimentally realized wurtzite, zincblende and rocksalt ZnO 
are the top occurring structures with the relative frequencies in the 0.04 - 0.11 range well separated from the rest. In the case of SnO$_2$, 
the relative frequencies are all below 0.02. The top two occurring are the ground-state rutile and the anatase structures. The $\alpha$-PbO$_2$ 
is the structure type with the highest occurrence among those that have their total energy close to, and volume smaller than the rutile. 
These results elucidate why is it relatively easy to push SnO$_2$ outside of the rutile structure using pressure (relatively small basin of attraction) 
and why it undergoes a series of phase transitions (all other structures have small basins), and why does $\alpha$-PbO$_2$ occur 
upon releasing the pressure.

{\it RSL Sampling.} The details of the RSL structures generation are shown in the upper panel of Fig.~\ref{fig2}. 
It is a modification of the Ab Initio Random Structure Searching method \cite{pickard_JPCM:2011} and similarly starts with the random choice 
of unit cell parameters $a, b, c, \alpha, \beta, \gamma$. In the second step the cation and anion grids are constructed in the following 
way. First, a transformation to the fractional (crystal) coordinates is performed to provide a cubic-like representation 
of the unit cell. Then, a reciprocal lattice vector ${\mathbf G} = n_1 {\mathbf g}_1 + n_2 {\mathbf g}_2 + n_3 {\mathbf g}_3$ with random $n_1,n_2,n_3$ 
is constructed. ${\mathbf G}$ defines a plane wave $cos({\mathbf G}{\mathbf r})$ and an associated superlattice. The two grids are constructed by 
discretizing the planes corresponding to the minima (cation grid) and the maxima (anion grid) of the plane wave. In the third step the ions are distributed 
over the two grids. To ensure homogeneous distribution and that that no two ions of the same kind are too close, the probability distribution is constructed 
by placing a gaussian centered at each occupied grid point. The next ion is then placed on a grid point chosen randomly among those that have low 
probability. Finally, the structure is converted from fractional back into the real coordinates and the scaling factor is adjusted such that the minimal distance 
between any two atoms is larger than a certain threshold. The example of an RSL structure of ZnO shown in the Supplemental Material clearly displays 
the dominant cation-anion coordination.

A total of 6000 RSL structures are generated (2000 per system) with the following parameters: $a, b$, and $c$ randomly chosen between 0.6 and 1.4 
(in units of {\it scale}); $\alpha, \beta$, and $\gamma$ random in the $(30^\circ, 160^\circ)$ range;  $n_1,n_2$, and $n_3$ also random between 4 and 10, 
the range which ensures that sufficient, but not too large, number of planes in the unit cell; and the {\it scale} is adjusted such that the 
shortest distance between the atoms is not shorter than 1.8 {\AA}. Different unit cell sizes are sampled by creating the RSL structures with one through 20 
formula units and 100 RSLs per size. Alternative would be to fix the cell size, but this would bias the sampling
only to structures with sizes compatible with the chosen one.  

{\it DFT Calculations.} Full relaxations, including volume, cell shape and atomic positions, are performed on all RSL structures. 
This is done by employing standard DFT approach \cite{vladan_PRB:2012} with the PBE form of the exchange-correlation functional \cite{perdew_PRL:1996}
and the projector augmented wave (PAW) method \cite{bloechl_PRB:1994} as implemented in the VASP code \cite{kresse_CMS:1996}.
The employed numerical setup ($\mathbf k$-points, various cutoffs) results in absolute total energies that are converged to within 3 meV/atom. 
All relaxations (volume, shape, atomic positions) are conducted using the conjugate gradient algorithm \cite{press:2007}. Construction of the workflows, 
management of large number of calculations, and the analysis of results is carried out using the {\it pylada} software 
package \cite{pylada}.

{\it Structure sorting.} Sorting of the resulting, DFT-relaxed, structures into the classes of equivalence is done based on four criteria. Two structures are 
considered equivalent if: (1) their total energies are within 10 meV/atom, (2) their space groups match, 
(3) their volumes per atom are within 0.5 \%, and (4) the coordination of atoms up to the 4th neighbor is the same. After extensive testing these four 
criteria were proven robust in establishing equivalence between the structures.  

{\it Results.} The RSL sampling procedure combined with DFT relaxations results in: 904 distinct structure types (classes of equivalence) for MgO, 1306 for 
ZnO and 1740 for SnO$_2$. 
The lower panel of Fig.~\ref{fig2} shows for all three systems the number of distinct structures within an energy window $\Delta E = 50, 75, 100, 125$, and 
$150$ meV/atom above the ground state as a function of the total number of structures used in sampling. Two observations can be made. First, there are 
clear differences between the PESs of MgO, ZnO and SnO$_2$, evidenced by a different number of distinct structures present in the corresponding $\Delta E$. 
Second, for all three systems the number of distinct structures grows monotonically with little, or no sign of convergence. This implies that even within a 
relatively narrow energy range a very large number of polymorphs can potentially exist. 

Fortunately however, a vast majority of the structures are actually irrelevant for their very low probability (frequency) of occurrence. As shown in 
Fig.~\ref{fig3}, except for a relatively small number of structures that occur more frequently, nearly all of the polymorphs resulting
from the sampling procedure occur only once. Moreover, if the frequency of occurrence is represented as a function of the total number of structures, much like 
in the lower panel of Fig.~\ref{fig2}, the extrapolation to the infinite number of structures would imply zero probabilities for all of these (see Supplemental Material). 
All of the top occurring structures have their frequencies of occurrence fairly converged implying
the finite probabilities of occurrence. All of the experimentally realized polymorphs appear among the top occurring structures.

In case of MgO, the top occurring is the rocksalt $Fm$-$3m$ structure with the relative frequency of occurrence of 0.437 far above 
any other structure. It is followed by the zincblende $F$-$43m$, and a four atom $P6_3/mmc$ structure with the relative frequencies 
of 0.0275 and 0.016, respectively. The occurrence of all other structures is below 0.01. It is not a surprise then that the rocksalt is the only experimentally 
realized MgO structure. Note that these arguments do not mean that realizing other structures is entirely impossible, but that
creating experimental conditions to do so might be much more challenging. 

The RSL sampling of ZnO polymorphs results in the distribution of the relative frequencies such that four distinct structures occur more frequently than the 
others. These are: the ground state wurtzite $P6_3mc$, zincblende $F$-$43m$, tetragonal $I4mm$, and the rocksalt 
$Fm$-$3m$. Phonon calculations reveal  that the tetragonal $I4mm$ shown in grey in Fig.~\ref{fig3} is dynamically unstable. The 
remaining three are exactly the three known, experimentally realized, polymorphs of ZnO \cite{icsd, ashrafi_APL:2000, decremps_APL:2002}. 
The ZnO results support directly the  previous discussion about the energy above the ground state and its
inadequacy to judge the realizability of different polymorphs. Namely, as shown in Fig.~\ref{fig3} there is a large number of distinct structures that 
appear inside the $\sim$150 meV/atom energy window between the ground state wurtzite and the rocksalt structure. A significant fraction of 
these have their volumes smaller than wurtzite. It is however, the rocksalt that is realized by applying the pressure despite its relatively high energy.
Based on these results it can be argued that the realization of the rocksalt phase is due to the relatively large volume of configuration of space 
occupied by the rocksalt structure.  Of course, for the polymorph to be metastable the kinetic barriers need to be sufficiently high to provide the 
conditions for the kinetic trapping, which is experimental fact for both zincblende and rocksalt ZnO \cite{icsd, ashrafi_APL:2000, decremps_APL:2002}. 
However, it is the volume of configuration space that is comes first in determining the list of candidate structures with higher likelihood to be 
experimentally realized. The magnitude of the kinetic barriers will then determine which of {\it these} structures will actually be metastable.

The results for SnO$_2$ further support this discussion. Two most frequent structures in the RSL sampling are the ground state rutile 
$P4_2mnm$ and the anatase $I4_1amd$ structure. As indicated in Fig.~\ref{fig3}, cubic $Fm$-$3m$ structure is dynamically unstable, while 
the fourth most frequent structure $R$-$3m$ is $\sim$200 meV/atom above rutile. The metastable $Pbcn$ structure ($\alpha$-PbO$_2$ type), 
that is observed in high pressure experiments, is the most frequent among the structures that have their energy close to, and volumes smaller 
than the rutile phase. This explains why is $Pbcn$ SnO$_2$ typically observed upon releasing the pressure. Interestingly, the anatase SnO$_2$ 
predicted here to be among the most probable structures has so far not been realized experimentally. It has the volume significantly larger than 
the rutile and, contrary to TiO$_2$ which has the rutile and anatase polymorphs nearly degenerate, has the energy by more than 50 meV/atom 
above rutile SnO$_2$. Therefore, the absence of anatase SnO$_2$ is likely due to the challenges associated to achieving 
experimental conditions to fulfill both of these conditions simultaneously (expanded volume and higher energy).

Finally, it is important to note that the actual values of the frequencies of occurrence do depend on the approach to sample different supercell sizes. 
As already mentioned, the supercells containing between 1 -- 20 f.u. are sampled homogeneously in this work. The reason is to avoid biasing of the 
sampling only to sizes compatible with some fixed supercell size. Tests of different ways to sample the system size  (fixed versus variable) show that 
the list of the top occurring structures that have their frequencies of occurrence converged with respect to the total number of structures does not 
depend on the sampling procedure.
%
%

In conclusion, this paper demonstrates that the random superlattice (RSL) structure sampling followed by the DFT relaxations 
can be used to screen different polymorphs of ionic systems as well as to assess the likelihood for their experimental realization. 
It is shown that the key quantity in assessing the likelihood for experimental realization is the resulting frequency of occurrence, 
which measures (indirectly) the volume of configuration space occupied by a different structures. Application of the RSL sampling 
on MgO, ZnO, and SnO$_2$ reveals that the experimentally observed polymorphs are exclusively the ones with the highest frequency 
(probability) of occurrence explaining the physical reasons behind the experimental observations.  

\begin{acknowledgments}
%
The author is grateful to S. Lany and G. Ceder for inspiring and fruitful discussions.
This work was supported as part of the Center for the Next Generation of Materials by Design, an Energy Frontier 
Research Center funded by the U.S. Department of Energy, Office of Science, Basic Energy Sciences under 
Contract No. DE-AC36-08GO28308 to NREL.
The research was performed using computational resources sponsored by the 
Department of Energy's Office of Energy Efficiency and Renewable Energy and located at the National 
Renewable Energy Laboratory.

\end{acknowledgments}

%
%
%
\end{document}